\documentclass{article}
\usepackage{graphicx}

\vspace{-20ex}
\title{Introduction to Spin \& Lattice Models\\
in the Social Sciences}
\author{Kenton K. Yee${}^{\dag}$\\
~\\
{\small\sl Columbia University\/}\\
~\\
{\small http://papers.ssrn.com/sol3/cf\_dev/AbsByAuth.cfm?per\_id=240309 }}
\begin{document}
\bibliographystyle{bibstand}
\begin{titlepage}
\maketitle
\begin{abstract}
\baselineskip=20pt
In recent years, 
political and financial economists and other social scientists have begun
adopting spin and lattice models into their theoretical tool kit.
This article introduces examples of how these models are used, and points
to some state of the art references.  
For illustration, a simple dynamical model
of how legal rules evolve and propagate in
the Anglo-American court system is described.  
\end{abstract}
\vfill
\baselineskip=15pt
\noindent
${}^{\dag}$The author is 
pleased to acknowledge financial support from
the John M. Olin Academic Year Fellowship in Law and Economics
at Stanford Law School and the
Santa Fe Institute's Complex Systems Summer School.  
\thispagestyle{empty}
\end{titlepage}
\pagebreak
\parindent=10mm                

\baselineskip24pt
\setlength{\arraycolsep}{0pt}
\vskip25pt

\section{The Common Law as an Evolutionary System}
\label{sec:intro}

Almost two centuries ago, Justice
Oliver Wendell Holmes wrote ``The 
life of the law has not been knowledge: it has been experience.  The 
felt necessities of the time, ... avowed or unconscious, even the
prejudices which judges share with their fellow men, have had a good deal
more to do than the syllogism in determining the rules by which men [are]
governed.''  Likewise, the German legal scholar
Friedrich Karl von Savigny wrote that ``All law... is 
first developed by custom and [conventional morality],
next by jurisprudence,---everywhere, therefore, by internal silently-operating
powers, not by the arbitrary will of a law-giver.'' 

Holmes' ``felt necessities'' and von Savigny's ``internal silently-operating
powers'' have not gone unnoticed by modern scholars.  Modern 
legal scholars, 
ranging from Grant Gilmore to Lawrence Friedman,
have observed that changes in law at
various times are ``rapid and violent,''
``clean and swift,'' and ``highly fluctuating'' (Yee 2001).  In
current events, a growing branch of intellectual property,
``cyberspace law,'' has emerged almost overnight
in response to technological innovations
and business needs.  Moreover, capital markets
anticipate such changes in a rational way (Yee 2005, 2006).

Hence, it comes as no surprise that
Holmes' ``felt necessities'' and 
von Savigny's ``internal silently-operating
powers'' drive the basis of an evolutionary framework for understanding
the developement the 
common law (Priest 1977; Rubin 1977).  The idea is based
on Darwinian natural selection: efficient
legal rules survive while inefficient rules are culled out by
litigation.  In the long run, the common law contains only rules
that survive legal and political challenges.  While this paradigm
is appealing, it has not yielded empirically refutable predictions 
which would seriously test the model.

This article introduces a dynamical model of common law
evolution originally described in Yee (2001).  According to this model, 
evolution leaves an inevitable trail of paleontological
footprints which may be 
sitting in the Westlaw dunes
awaiting empirical identification by legal excavators.  

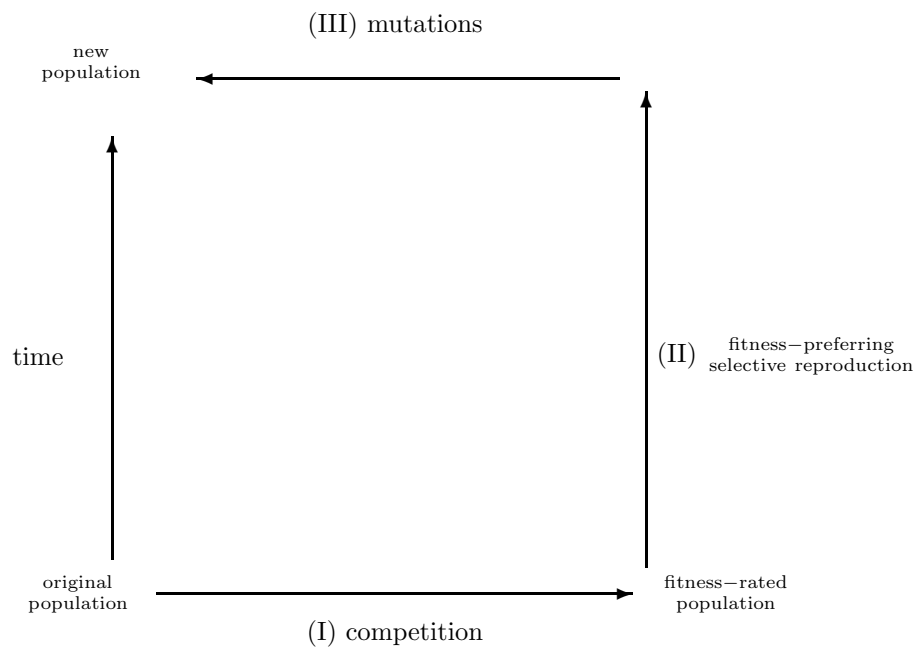
\begin{figure}[bpt]
\begin{picture}(350,240)(-50,0)
\thicklines
\put(0,140){\makebox(0,0){time}}
\put(135,265){\makebox(0,0){(III) mutations}}
\put(135,35){\makebox(0,0){(I) competition}}
\put(283,140){\makebox(0,0){(II) ${{\rm fitness-preferring\/} \atop 
{\rm selective~reproduction\/}}$}}
\put(15,50){\makebox(0,0){${{\rm original\/}\atop {\rm population\/}}$}}
\put(20,250){\makebox(0,0){${{\rm new\/}\atop {\rm population\/}}$}}
\put(260,50){\makebox(0,0){${{\rm fitness-rated\/}\atop {\rm population\/}}$}}
\put(45,50){\vector(1,0){180}}
\put(230,60){\vector(0,1){180}}
\put(28,63){\vector(0,1){160}}
\put(220,245){\vector(-1,0){160}}
\end{picture}
\caption{A Darwinian evolutionary system evolves in time as follows.  At
any time step $t$, there is 
a population of individuals.  Mutual competition
establishes 
a fitness rating for each individual.  The population then undergoes
natural selection based on the fitness rating, 
selective reproduction, and 
(usually random) mutations.  This 
process yields a new population of individuals at time $t+1$, the
subsequent time step.}
\label{fig:evtree}
\end{figure}

Biological evolution starts with
a collection of genes (random degrees of freedom)
created by chance chemistry in the earth's primordial atmosphere.  These 
early genes competed against each other to survive and 
replicate.  Ultimately, the biological structures
we see today, including their inorganic 
byproducts such as computer software or the price of wheat futures
emerged from this competition.  Emergence 
is not imposed exogenously.  Rather,
structures emerge inevitably from 
natural selection, which in turn is an 
inherent consequence of competitive interactions and 
selective reproduction. 

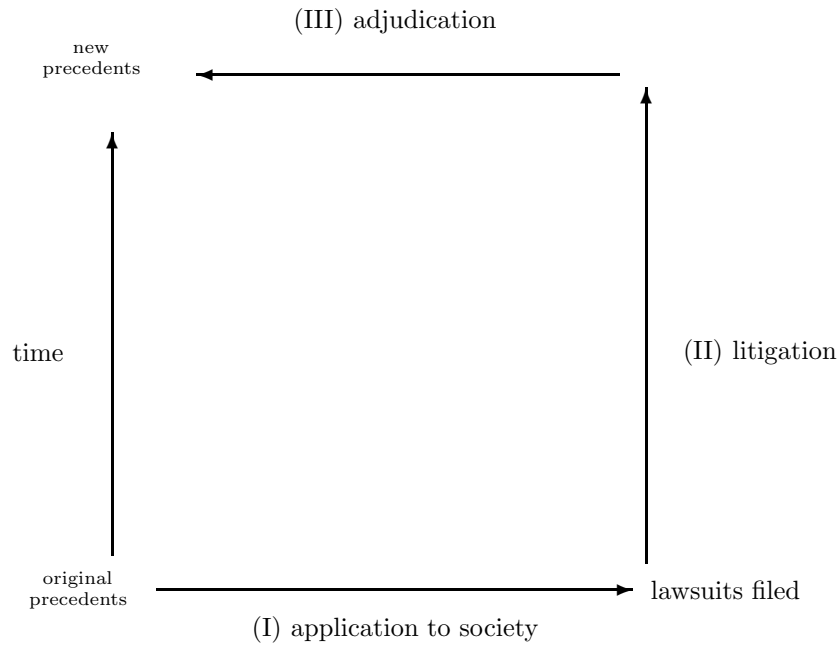
\begin{figure}[tbp]
\begin{picture}(350,240)(-50,0)
\thicklines
\put(0,140){\makebox(0,0){time}}
\put(135,265){\makebox(0,0){(III) adjudication}}
\put(273,140){\makebox(0,0){(II) litigation}}
\put(135,35){\makebox(0,0){(I) application to society}}
\put(15,50){\makebox(0,0){${{\rm original\/}\atop {\rm precedents\/}}$}}
\put(20,250){\makebox(0,0){${{\rm new\/}\atop {\rm precedents\/}}$}}
\put(260,50){\makebox(0,0){${{\rm lawsuits~filed\/}}$}}
\put(45,50){\vector(1,0){180}}
\put(230,60){\vector(0,1){180}}
\put(28,63){\vector(0,1){160}}
\put(220,245){\vector(-1,0){160}}
\end{picture}
\caption{Interpretation of Figure~\ref{fig:evtree} when the evolving 
system is 
the common law (or any precedent-based legal regime).  At
any time step $t$ is
a collection of laws or precedents.  As these precedents
are applied to regulate social and economic activity,
disputes arise concerning 
their meaning and social 
desirability.  Accordingly, lawsuits (or legislative challenges) 
seeking to overturn
the precedents in controversy are filed.  One
precedent is more legally ``fit'' than another if
it suffers fewer such challenges.  Copies 
of the unchallenged (and hence the fittest) 
precedents survive intact to reign at subsequent time step 
$t+1$.  
The challenged precedents, facing judicial modification or
termination with each litigation, have a decreased
chance of unaltered survival.  Moreover, all
precedents (challenged and unchallenged) 
are subject to random mutations 
stemming from external social
pressures which may alter their 
meaning or applicability at any time.  Thus at each new time step 
is an evolving common law comprised of 
precedents which are
selectively reproduced and mutated replicas of their ancestors.
} 
\label{fig:litigtree}
\end{figure}

Natural selection and 
evolution does not only occur in a biological setting.  Any
system is evolutionary if it has 
the following four ingredients: set of degrees of freedom or individuals, 
each ranked by a quality or behavior-based fitness criterion, a 
selection process based on the fitness ranking, and a 
mechanism for introducing (usually random) 
mutations to periodically interject diversity into the 
reproductive process.  

The common law 
has these four ingredients.  
As depicted in Figure~\ref{fig:litigtree},
precedents of the common law are the ``individuals'' undergoing
Darwinian evolution.  A precedent's ``economic efficiency'' 
is the degree to which it
balances between
legal and political forces.  Economic efficiency plays the role
of Darwinian fitness.  The 
fitness-preferring selection process is provided by the premise that
inefficient precedents are subject to more
challenges (either in the courts or in the legislature) 
and hence are more subject to alteration than efficient ones.  Mutations
are introduced by the volatile nature of litigation or legislation.

	As pointed out by many authors, notably 
Priest (1977) and Rubin (1977), 
common law precedents tend to evolve towards economic efficiency 
because inefficient precedents are selectively
challenged and ultimately culled away as 
judges eventually hit on more efficient and hence less-challenged
doctrines even if only by
random trial and error.  In this view, the common law is a 
market of doctrines, and a law suit is a bid on a specific doctrine.  
Intensive litigation moves this market
towards efficiency whether or not judges
consciously choose efficiency because inefficient doctrines will
be intensively relitigated until they are efficient. 

	While this market view is a compelling
premise, it is not all there is
to the story.  Evolution 
is more than a push towards improvement.  Biological
evolution has provided a
rich history of paleontological footprints:
punctuated equilibria, Zipf's Law, 
and path-dependency.  Accordingly, taking 
the evolutionary paradigm seriously requires
considering
its dynamical properties---the 
paleontological footprints.  In Yee (2001) I 
propose a dynamical 
model (``CLM'') of common law evolution.  Cast as parsimoniously
as possible, CLM in its simplest guise is mathematically 
isomorphic to the Bak-Sneppen models (Bak and Sneppen
1993; see also Yee 1993) and yields interesting 
paleontology. As discussed, CLM exhibits
punctuated equilibrium, Zipf's Law, path dependency, and
a stochastic (but statistically robust) 
form of efficiency I shall refer to as ``smeared'' economic
efficiency.

\section{The Common Law Model (``CLM'')}
\label{sec:model}

Economic efficiency of a precedent can change either because
($\alpha$) the precedent is altered as a result of a court decision or
legislation; or its ($\beta$) legal environment or
($\gamma$) social context
changes while the precedent itself remains constant.  ($\beta$) recognizes
that changes in related laws may induce a change in
the economic efficiency of
an unaltered precedent.  For instance, a modification of traffic laws may
distort the economic efficiency of a prevailing
``reasonable man'' standard in torts without any direct
change to the standard itself.  Likewise, ($\gamma$) says that
social or cultural changes, perhaps driven by
technological innovations, may induce the economic efficiency of
a law to change without any changes to the law itself.
In this Introduction, I shall assume that changes in
economic efficiency are due entirely to ($\alpha$) and ($\beta$), 
not ($\gamma$).

With this caveat in mind,
my CLM has just two ingredients:
\begin{itemize}
\item a set of $N\ge 3$ precedents labeled by integers $i\in
I \equiv \{1,2,\cdots,N\}$;
\item a real-valued economic fitness
measure $e:I\longmapsto [0,1]$ where
$e=0$ represents the worst and $e=1$ the best
possible economic efficiency level.
\end{itemize}
CLM evolves in time according to the following three rules:
\begin{itemize}
\item (I) the precedent $i$ litigated at each time step is the
one with the smallest efficiency value $e(i)$;
\item (II) litigation of
precedent $i$ results in a court ruling which
potentially alters not only $i$ but also $i$'s
neighboring precedents\footnote{I
assume that precedent space $I$ wraps around so that
$i=N$ and $i=1$ are next door neighbors.  This wrap-around
assumption has negligible
effect on my results when $N$ is sufficiently
large (e.g. if
$N >> 10$).  In real life, the common law arguably has
thousands of precedents.} $i + 1$ and $i-1$;
\item (III) litigation outcomes are random in
efficiency $e$, that is courts don't strive to optimize efficiency.  In
view of Rule (II), this means
a litigation of precedent $i$ results in
$e(i)$, $e(i-1)$, and $e(i+1)$ each being assigned a new
random value in the unit interval.
\end{itemize}

These three rules drive CLM.  As I shall describe,
this minimalistic set of rules yield
a wealth of interesting evolutionary
implications for common law behavior.  Rule (I) represents
the Rubin-Priest conjecture of
selective litigation of the most inefficient laws.  Rule (II)
is motivated by ($\beta$), that is precedents
do not reign in a vacuum but are interpreted within its legal context
at each point in time.  A precedent's meaning and application---and
accordingly its economic efficiency---depends on the meaning and application
of its cousins in legal logic and, ultimately,
on the entire common law.  In Rule (II), the multifaceted
web of legal relationships between different precedents is represented by
the bold simplification that
changes in precedent $i$ directly influences only its next-door neighbors.

Rule (III) assumes
that litigation results are entirely
random with respect to economic efficiency.  While this assumption
was selected for its minimalistic nature, it
is not as detached from real life as some practitioners
are apt to believe.  In addition to the everchanging
mix of judicial philosophies and idiosyncrasies flowing in and out of
the judicial system, judges also
make their share of mistakes.  As it is,
even the Supreme Court's decisions
are demonstrably random in some contexts, such as
in at least one area of securities regulation.

A simplified toy model helps elucidate
CLM.  For the toy model,
assume efficiency of each precedent can take on
only three values High, Medium, and Low, and litigation of
one precedent bears no consequences for its neighbors.  While
these assumptions ignore critical elements of CLM---notably Rule
II, the interprecedent interaction rule, it serves as a starting point to
develop intuition.  In the earlier rounds of the
toy model, one third of
the precedents will be Low, and they will be
litigated first.  In each trial, two times out of three the judge
will replace the Low precedent with one of a higher score.
After enough rounds of litigation, all the Low
precedents will be eliminated, if only for one round.  When
all the Low precedents
are eliminated, a random Medium precedent will be litigated in the
subsequent time step.  There are three possible outcomes.  If
the judge offers a High result, lawyers will move on
to litigate other Medium precedents.  If
the judge offers a Medium ruling, the common law has not deteriorated.
If the judge offers a Low ruling, the Low precedent will be immediately
relitigated, again and again if necessary,
until it is restored to a Medium or High efficiency.
Repeated, instantaneous
relitigation of a Low precedent generates a
``litigation cluster'' in the toy model.

In the toy model, the
trend is clear: Low precedents are relitigated
relentlessly until they become extinct.  Subsequently,
Medium precedents are culled until only High precedents remain.  Except
for short-lived litigation clusters,\footnote{When all the other precedents
are High, a litigated precedent has
a two thirds chance of being immediately relitigated.}
only High precedents survive in the
long run.  Therefore, the long run equilibrium state of the
toy model is perfectly efficient (except for small litigation clusters).
In equilibrium, High pecedents live at the ``edge'' of litigation; that is,
as the only surviving precedents,
they are randomly exposed to litigation and, when litigated,
are relentlessly rehashed until restored to High efficiency.

The critical difference between 
the toy  model and CLM is due to Rule II.  Rule II, 
interprecedent interactions, in
CLM is responsible for a rich variety of effects in CLM.
Because of interprecedent interactions, High precedents
are as vulnerable to litigation as their Low or Medium neighbors are.
In CLM, High neighbors of a Low precedent
are subject to being pulled into the litigation undertow whenever
the Low neighbor is litigated.  Since the undertow consists of
up to three precedents, and only one precedent can be litigated
at each step, the efficiency distribution does not collapse into
a flat $e=1$ peak.  Instead,
litigation of a Low or Medium precedent creates a
lasting wave of litigation in the whole surrounding neighborhood.  This
enhances litigation clusters, creates valleys of inefficient
litigation, and smears out the efficiency distribution.

Real life precedents, like in CLM and contrary to the toy model,
do not act independently of
the remaining body of law.  Precedents depend on or
complement and reinforce one another in a dynamic Charlotte's
web of legal and logical entanglements.  It
is precisely the presence of 
interprecedent entanglements
springing from
Rule II\footnote{While one
can extend Rule II by introducing next-to-nearest neighbor
interprecedent couplings,
in keeping with minimalism I shall focus exclusively on
nearest-neighbor couplings.  With only nearest-neighbor couplings,
CLM is similar to a physics model (Bak and Sneppen 1993).} that
differentiates CLM from
CK.  And, as shown in Yee (2001),
it is precisely the dynamics of these entanglements
which leave the paleontological footprints of
punctuated equilibria, Zipf's Law, path dependency,
and smeared economic efficiency.

\begin{figure}[tbp]
\centering
\includegraphics[width=5.2in]{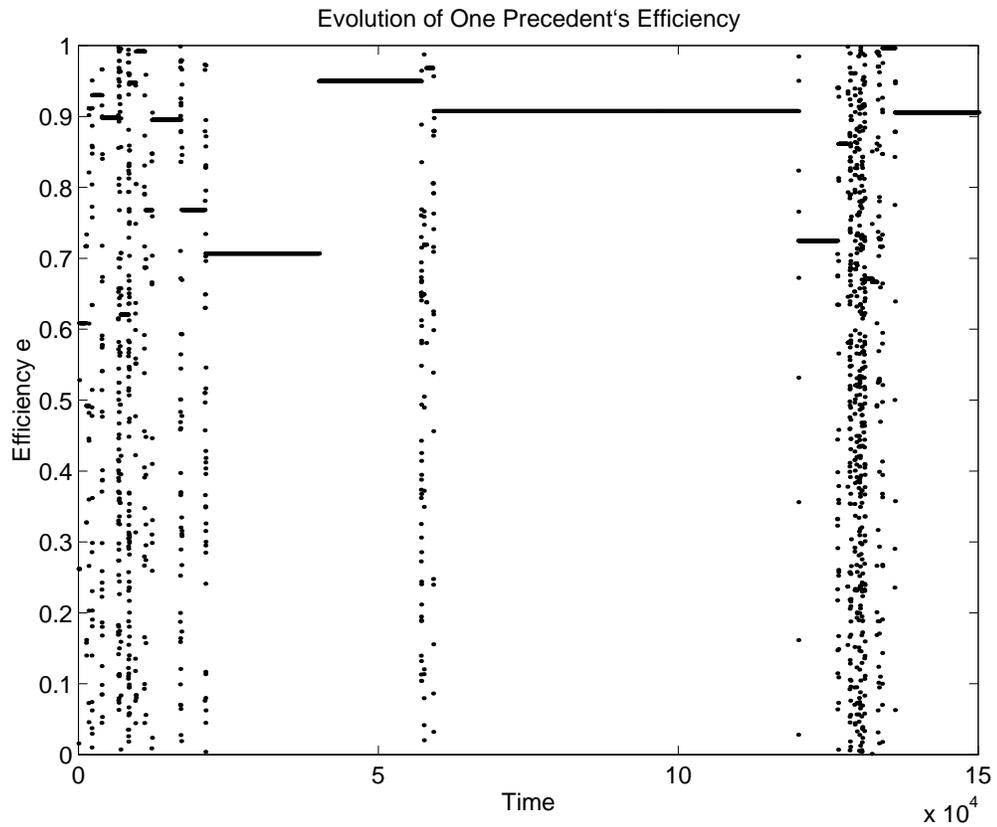}
\caption{Evolution of a typical precedent's efficiency value
as a function of time in CLM.  While equilibrium is reached some time before
the $5\times 10^4$th time step, the precedent continues to evolve and
fluctuate in efficiency.  Thus, equilibrium is not static:
while certain collective properties of
the system stablize at equilibrium, individual precedents continue to
evolve and mutate.
}
\label{fig6}
\end{figure}

For instance, as depicted in Figure~\ref{fig6},
a law exhibits
rapid violent change interspersed with long periods of quiet.  This
feature occurs endogenously.  In CLM,
spurts of intense litigation for each precedent are
interspersed with irregular periods of inactivity.  Figure~\ref{fig6} depicts
the efficiency of a typical precedent as a function of time.


\begin{thebibliography}{10}
\baselineskip14pt

\bibitem{baksneppen}
Bak, P., Sneppen, K., ``Punctuated Equilibrium and Criticality
in a Simple Model of Evolution,'' (1993) 24 {\sl Physical Review
Letters\/} 4083.

\bibitem{priest}
Priest, G., ``The Common Law Process and the Selection of
Efficient Rules,'' (1977) 6 {\sl J. Legal Stud\/}. 65.

\bibitem{rubin} 
Rubin, P., ``Why is the Common Law Efficient?'' 
(1977) 6 {\sl J. Legal Stud\/}. 51.

\bibitem{yee}
Yee, K., ``Central Charge of the Parallelogram Lattice
Strong Coupling Schwinger Model,''
(1993) D47 {\sl Physical Review\/} 1719-1722.

\bibitem{location}
Yee, K. (2006) ``Earnings Quality and the Equity Risk Premium: A Bemchmark Model,''
{\sl Contemporary Accounting Research\/}, forthcoming; 
PDF file available at\\
{\it http://papers.ssrn.com/sol3/papers.cfm?abstract\_id=846546}

\bibitem{adapt}
Yee, K. (2005) ``Aggregation, Dividend Irrelevancy, and 
Earnings-Value Relations,''
22.2 {\sl Contemporary Accounting Research\/}
453-480; PDF file available at\\
{\it http://papers.ssrn.com/sol3/papers.cfm?abstract\_id=667781}

\bibitem{yeeb}
Yee, K. (2001) ``Common Law Efficiency under Haphazard Adjudication,'' 
{\sl Columbia University\/} working paper, PDF file available at\\
{\it http://papers.ssrn.com/sol3/papers.cfm?abstract\_id=270593}
\end{thebibliography}
\end{document}